\documentstyle[preprint,aps,eqsecnum,amsfonts]{revtex}
\input epsf
\def\upA{\uparrow}
\def\dnA{\downarrow}
\def\llangle{\langle\!\langle}
\def\rrangle{\rangle\!\rangle}
\def\F{\Phi }
\def\FF{\widehat{\F}}
\def\LLL{lowest Landau level\ }
\def\LLLd{lowest Landau level}

\def\bbox#1{#1}
\def\Laugh{\F_{\text{LN}}[\bbox{x}]}
\def\text#1{{\rm #1}}
\def\textit#1{{\it #1}}
\def\CB{CB\ }
\def\CBs{composite bosons}
\def\e{\text{e}}
\def\i{\text{i}}
\def\dx{\text{d}^{2}x}
\def\dy{\text{d}^{2}y}
\def\dq{\text{d}^{2}q}
\def\spin{\text{spin}}
\def\bpmatrix{\left(\begin{array}{c}}
\def\epmatrix{\end{array}\right)}
\def\sky{\text{sky}}
\def\a{{\cal A}}
\def\B{{\cal B}}
\def\H{{\cal H}}
\def\P{{\cal P}}
\def\OO{\widehat{O}}
\def\Wed#1#2{\mg{#1\!\wedge\!#2\over2}}
\def\mg{\ell_{B}^{2}}

\begin{document}
\tightenlines 
\title{
Skyrmions and Quantum Hall Ferromagnets in
Improved Composite-Boson Theory}
\author{Zyun F. Ezawa and Kenichi Sasaki}
\author{Department of Physics, Tohoku University, Sendai 980-8578, Japan}
\maketitle
\begin{abstract}
An improved composite-boson theory of quantum Hall ferromagnets is proposed.  
It is tightly related with the microscopic wave-function theory.  The 
characteristic feature is that the field operator describes solely the 
physical degrees of freedom representing the deviation from the ground state.  
It presents a powerful tool to analyze excited states within the \LLLd.  
Excitations include a Goldstone mode and nonlocal topological solitons.  
Solitons are vortices and Skyrmions carrying the U(1) and SU(2) topological 
charges, respectively.  Their classical configurations are derived from their 
microscopic wave functions.  The activation energy of one Skyrmion is 
estimated, which explains experimental data remarkably well.
\end{abstract}
\sloppy

\section{Introduction}

The quantum Hall (QH) effect \cite{FQHEbook} is a remarkable macroscopic 
quantum phenomenon observed in the two-dimensional electron system at low 
temperature $T$ and in strong magnetic field $B$.  The Hall conductivity is 
quantized with extreme accuracy and develops a series of plateaux at magic 
values of the filling factor $\nu =2\pi \hbar \rho _{0}/eB$.  Here, $\rho _{0}$ is the average electron 
density.  The QH effect comes from the realization of an incompressible 
quantum fluid.  The composite-boson (CB) picture 
\cite{LGCSx,ReadA,RajaramanCB} and the composite-fermion picture \cite{JainCF,OtherCF} 
have proved to be quite useful to understand all essential aspects of QH 
effects.  Electrons may condense into an incompressible quantum Hall liquid as 
composite bosons.  The QH state is such a condensate of composite bosons, 
where quasiparticles are vortices \cite{LaughlinA}.  

When the spin degree of freedom is incorporated, a quantum coherence 
develops spontaneously provided the Zeeman effect is sufficiently small, 
turning the QH system into a QH ferromagnet.  New excitations are Skyrmions.  
Skyrmions were initially considered as solutions of the effective nonlinear 
sigma model \cite{SkyrmQH}, and later studied also in a Hartree-Fock approximation 
\cite{SkyrmHartFock}.  Their existence has been confirmed experimentally 
\cite{SkyExp,SkyExpEneA}.  

In this paper we present a field-theoretical formulation of QH 
ferromagnets and Skyrmions based on an improved composite-boson (CB) theory 
\cite{EzaICBa}, proposed based on a suggestion due to Girvin \cite{FQHEbook}, Read 
\cite{ReadA} and Rajaraman et al. \cite{RajaramanCB}.  In this scheme, the field 
operator describes solely the physical degrees of freedom representing the 
deviation from the ground state, and the semiclassical property of excitations 
is determined directly by their microscopic wave functions.  This clarifies 
the relation between the microscopic wave function and the classical Skyrmion, 
which is missing in the effective theory \cite{SkyrmQH}.  It is explicitly shown 
that the Skyrmion excitation is reduced to the vortex excitation in the 
small-size limit.  We estimate its activation energy, which accounts for the 
observed data due to Schmeller et al.\cite{SkyExpEneA} remarkably well provided 
certain physical assumptions are made.  Thus, our field-theoretical approach 
presents a new insight into the Skyrmion physics.  

This paper is composed as follows.  
In Section \ref{ReviewCB} we give an overview of the improved \CB theory.
In Section \ref{SecBoson} \CB fields are defined.  First, as in the standard 
\CB theory \cite{LGCSx}, we attach an odd number of flux quanta to an electron by 
way of a singular phase transformation.  We call the resulting electron-flux 
composite the \textit{bare composite boson}.  In order to soften the 
singularity brought in, we dress it with a cloud of an effective magnetic 
field that bare \CBs\ feel.  The resulting object turns out to be the 
\textit{dressed composite boson}.  
In Section \ref{SecQHS} the relation is established between the electron wave 
function and the \CB wave function.  We also verify that the ground state is 
given by the Laughlin state within the semiclassical approximation.  
In Section \ref{SecTE} we make a semiclassical analysis of vortex 
excitations.  In Section \ref{SecQHFerro} we include the spin degree of 
freedom and show that a quantum coherence develops spontaneously when the 
Zeeman effect is sufficiently small.  
In Section \ref{SecSpinTE} we discuss vortex and Skyrmion excitations in the 
QH ferromagnet.  The Skyrmion classical configuration is also derived directly 
from its microscopic wave function.  We evaluate the excitation energy of one 
Skyrmion and compare it with experimental data.  Throughout the paper we use 
the natural units $\hbar =c=1$.
\section{Overview}\label{ReviewCB}

We start with a review of the improved CB theory applyied to the 
spin-frozen QH system \cite{EzaICBa}.  We denote the electron field by $\psi (\bbox{x})$ and 
its position by the complex coordinate normalized as $z=(x+\i y)/2\ell _{B}$.  Any 
state $|\F\rangle $ in the \LLL at the filling factor $\nu =1/m$ ($m$ odd) is 
represented by the wave function,
\begin{equation}
\F[\bbox{x}] \equiv  \langle 0|\psi (\bbox{x}_{1})\cdots \psi (\bbox{x}_{N})|\F\rangle  = \omega [z]\Laugh ,
\label{WaveElect}
\end{equation}
where
\begin{equation}
\Laugh=\prod _{r<s}(z_{r}-z_{s})^{m}\exp[-\sum _{r=1}^{N}|z_{r}|^{2}]
\label{LaughWave}
\end{equation}
is the Laughlin function, and $\omega [z]\equiv \omega (z_{1},z_{2},\cdots ,z_{N})$ is an analytic function 
symmetric in all $N$ variables.  The mapping from the fermionic wave function 
$\F[\bbox{x}]$ to the bosonic function $\omega [z]$ defines a bosonization.  We call the 
underlying boson the \textit{dressed composite boson} and denote its field 
operator by $\varphi (\bbox{x})$.  The field operator turns out to be the one considered 
first by Read \cite{ReadA} and revived recently by Rajaraman et al. 
\cite{RajaramanCB}.  We derive that
\begin{equation}
\F_{\varphi }[\bbox{x}] \equiv  \langle 0|\varphi (\bbox{x}_{1})\cdots \varphi (\bbox{x}_{N})|\F\rangle  = \omega [z].
\label{WaveFunctDress}
\end{equation}
The Laughlin state is represented by $\F_{\varphi }[\bbox{x}]=1$.  
\subsection{Vortices}

A typical vortex state is represented by $\F_{\varphi }[\bbox{x}]=\prod _{r}^{N}z_{r}$.  When the wave 
function is factorizable, $\omega [z]=\prod _{r}^{N}\omega (z_{r})$, as in this example, we obtain a 
semiclassical equation,
\begin{equation}
\langle \varphi (\bbox{x})\rangle =\omega (z) .
\label{SemiClassConst}
\end{equation}
This is a highly nontrivial constraint since it connects directly the 
classical field $\langle \varphi (\bbox{x})\rangle $ to the microscopic wave function $\omega (z)$.  It dictates 
how the electron density $\rho (\bbox{x})$ is modulated around the topological 
excitation.

The topological charge density $Q(\bbox{x})$ is determined in terms of the 
classical field.  It is the vorticity, $Q(\bbox{x})\equiv Q^{V}(\bbox{x})$ given by
\begin{equation}
Q^{V}(\bbox{x})={1\over 2\pi \i}\varepsilon _{ij}\partial _{i}\partial _{j}\ln\langle \varphi (\bbox{x})\rangle  =\delta (\bbox{x}) ,
\label{Vorte}
\end{equation}
for a vortex sitting at the origin.  The topological soliton induces the 
density modulation according to the soliton equation,
\begin{equation}
{\nu \over 4\pi }\bbox{\nabla }^{2}\ln \rho (\bbox{x}) - \rho (\bbox{x}) + \rho _{0}= \nu Q(\bbox{x}) ,
\label{SolitEq}
\end{equation}
as follows from the semiclassical constraint (\ref{SemiClassConst}).  The 
topological soliton carries the electric charge $\nu qe$, where $q=\int d^{2}x Q(\bbox{x})$ is 
the topological charge.
\subsection{Skyrmions}

We next summarize the idea of an improved CB theory applied to QH 
ferromagnets.  We denote the spin component by the index $\alpha (=\upA ,\dnA )$.  Any state 
at $\nu =1/m$ is represented by the wave function similar to (\ref{WaveElect}).  Let 
us explicitly consider the case when the spinor component is factorizable,
\begin{equation}
\F^{\text{spin}}[\bbox{x}] = \prod _{r}\bpmatrix \omega ^{\upA }(z_{r})\\ \omega ^{\dnA }(z_{r}) \epmatrix_{\kern-1pt r} \Laugh.
\end{equation}
The ground state is given by 
\begin{equation}
\F^{\spin}_{\text{g}}[\bbox{x}] = \prod _{r}\bpmatrix1\\0\epmatrix_{\kern-1pt r}\Laugh.
\label{GrounSpin}
\end{equation}
From a set of two analytic functions $\omega ^{\alpha }(z)$ we construct the 
complex-projective field $\bbox{n}(\bbox{x})$ whose two components $n^{\alpha }(\bbox{x})$ are
\begin{equation}
n^{\alpha }(\bbox{x}) = {\omega ^{\alpha }(z)\over \sqrt {|\omega ^{\upA }(z)|^{2}+|\omega ^{\dnA }(z)|^{2}}}. 
\end{equation}
This represents the most general Skyrmion configuration \cite{Skyrmion} together 
with the asymptotic boundary condition $n^{\upA }=1$ and $n^{\dnA }=0$.  The normalized spin 
field, or the nonlinear sigma field, is defined by
\begin{equation}
s^{a}(\bbox{x})=\bbox{n}(\bbox{x})^{\dagger }\tau ^{a}\bbox{n}(\bbox{x}) ,
\label{SigmaField}
\end{equation}
with $\tau ^{a}$ the Pauli matrices.  The Skyrmion configuration is indexed by the 
Pontryagin number \cite{Skyrmion}, whose density is
\begin{equation}
Q^{P}(\bbox{x}) = {1\over 8\pi }\varepsilon _{abc}\varepsilon _{ij}s^{a}(\bbox{x})\partial ^{i}s^{b}(\bbox{x})\partial ^{j}s^{c}(\bbox{x}) .
\label{PontrNumbe}
\end{equation}
The Skyrmion excitation modulates not only the spin density but also the 
electron density.  The electron density modulation is determined by the same 
soliton equation as (\ref{SolitEq}), where $Q(\bbox{x})\equiv Q^{P}(\bbox{x})$ is the Pontryagin number 
density (\ref{PontrNumbe}).  

The simplest excitation is given by one Skyrmion with scale $\kappa $ sitting 
at $\bbox{x}=0$, whose wave function is \cite{Moon}
\begin{equation}
\F^{\spin}_{\sky}[\bbox{x}] = \prod _{r}\bpmatrix z_{r}\\ \kappa /2 \epmatrix_{\kern-1pt r} \Laugh.
\label{SimplSkyrm}
\end{equation}
The scale $\kappa $ is to be fixed dynamically to minimize the excitation energy.  
The Pontryagin number density (\ref{PontrNumbe}) is calculated for the simplest 
Skyrmion (\ref{SimplSkyrm}) as
\begin{equation}
Q^{P}(\bbox{x})= {1\over \pi } {(\kappa \ell _{B})^{2}\over [r^{2}+(\kappa \ell _{B})^{2}]^{2}} .
\end{equation}
We study the limit $\kappa \rightarrow 0$, where the Skyrmion wave function (\ref{SimplSkyrm}) is 
reduced to the vortex wave function in the spin-polarized ground state, and  
the Pontryagin number density is reduced to the vorticity (\ref{Vorte}), 
$Q^{P}(\bbox{x})\rightarrow \delta (\bbox{x})$, as $\kappa \rightarrow 0$.  Hence, the soliton equation (\ref{SolitEq}) yields a 
vortex configuration for the zero-size Skyrmion.  This makes clear the 
relation between the Skyrmion and the vortex in QH ferromagnets.  
\section{Bosonization}\label{SecBoson}

The microscopic Hamiltonian for spinless planar electrons in external 
magnetic field $(0,0,-B)$ is given by
\begin{mathletters}
\begin{eqnarray}
H&=&{1\over 2M}\int \dx \psi ^{\dagger }(\bbox{x})(P_{x}^{2}+P_{y}^{2})\psi (\bbox{x}) + H_{C}    \label{Hamil:A}\\
&=&{1\over 2M}\int \dx \psi ^{\dagger }(\bbox{x})(P_{x}-\i P_{y})(P_{x}+\i P_{y})\psi (\bbox{x}) + {N\over 2}\omega _{c} + H_{C} , \label{Hamil:B}
\end{eqnarray}
\end{mathletters}
where $\psi (\bbox{x})$ is the electron field; $P_{j}=-\i \partial _{j}+eA^{\text{ext}}_{j}$ is the covariant 
momentum with $A^{\text{ext}}_{j}={1\over 2}\varepsilon _{jk}x_{k}B$; $N$ is the electron number and $\omega _{c}$ 
is the cyclotron frequency.  The Coulomb interaction term is
\begin{equation}
H_{C} = {e^{2}\over 2\varepsilon }\int \dx\dy {\varrho  (\bbox{x})\varrho  (\bbox{y})\over |\bbox{x}-\bbox{y}|} , 
\label{CouloEnergPre}
\end{equation}
where $\varrho  (\bbox{x})\equiv \rho (\bbox{x})-\rho _{0}$ stands for the deviation of the electron density 
$\rho (\bbox{x})\equiv \psi ^{\dagger }(\bbox{x})\psi (\bbox{x})$ from its average value $\rho _{0}$.

The state $|\F\rangle $ in the \LLL obeys
\begin{equation}
(P_{x}+\i P_{y})\psi (\bbox{x})|\F\rangle =-{\i\over \ell _{B}}\biggl(z + {\partial \over \partial z^{*}}\biggr)\psi (\bbox{x})|\F\rangle  = 0 ,
\label{LLLcondiElect}
\end{equation}
upon which the kinetic Hamiltonian is trivial.  We call it the LLL condition.  
The generic solution of this equation yields the $N$-body wave function 
(\ref{WaveElect}) for the state $|\F\rangle $.  We are concerned about the state $|\F\rangle $ in 
the \LLLd.

The bosonization scheme was pioneered by Girvin and MacDonald 
\cite{LGCSx}.  We define the field $\phi (\bbox{x})$ by an operator phase transformation,
\begin{equation}
\phi (\bbox{x}) = \e^{-\i\Theta (\bbox{x})}\psi (\bbox{x}) ,
\label{NaiveField}
\end{equation}
where $\Theta (\bbox{x})$ is the phase field,
\begin{equation}
\Theta (\bbox{x}) = m\int \dy \theta (\bbox{x}-\bbox{y})\rho (\bbox{y}) ,
\label{PhaseTheta}
\end{equation}
with the azimuthal angle $\theta (\bbox{x}-\bbox{y})$.  When $m$ is an odd integer, we can prove 
that $\phi (\bbox{x})$ is a bosonic operator.  Let us call the underlying boson the bare 
composite boson.  It is a hardcore boson satisfying the exclusion principle, 
$\phi (\bbox{x})^{2}=0$.  The LLL condition (\ref{LLLcondiElect}) reads
\begin{equation}
(\check{P}_{x}+\i\check{P_{y}})\phi (\bbox{x})|\F\rangle =0 ,
\label{LLLcondiBare}
\end{equation}
where $\check{P_{j}}\equiv P_{j}+\partial _{j}\Theta (\bbox{x})$ is the covariant momentum for the bare \CB 
field.  By solving this condition, the $N$-body wave function is found to be 
$\F_{\phi }[\bbox{x}] \equiv  \omega [z]|\Laugh|$.  Namely, the standard bosonization is a mapping from 
the fermionic wave function $\F[\bbox{x}]$ to a bosonic wave function $\F_{\phi }[\bbox{x}]$,
\begin{equation}
\F[\bbox{x}] \mapsto  \F_{\phi }[\bbox{x}] \equiv  \omega [z]|\Laugh|. \label{NaiveBoson}
\end{equation}
The mapping is to attach $m$ Dirac flux quanta $m\phi _{D}$ to each electron by way 
of the phase transformation (\ref{NaiveField}), where $\phi _{D}=2\pi /e$.  Thus, the ground 
state of bare composite bosons is described by the modulus of the Laughlin 
function, $|\Laugh|$.  

We wish to introduce another \CB field $\varphi (\bbox{x})$ which makes the LLL 
condition as simple as possible.  We set
\begin{equation}
\varphi (\bbox{x}) = \e^{-\a(\bbox{x})-\i\Theta (\bbox{x})}\psi (\bbox{x}) ,
\label{DressField}
\end{equation}
together with an operator $\a(\bbox{x})$ to be determined later.  By substituting 
this into the LLL condition (\ref{LLLcondiElect}), provided $\a(\bbox{x})$ satisfies
\begin{equation}
\partial _{j}\a(\bbox{x}) = \varepsilon _{jk}\bigl\{\partial _{k}\Theta (\bbox{x}) + eA_{k}^{\text{ext}}(\bbox{x})\bigr\} = \varepsilon _{jk}\partial _{k}\Theta (\bbox{x}) - {1\over 2\ell _{B}^{2}}x_{j} ,
\label{EqForA}
\end{equation}
it is reduced to a simple formula,
\begin{equation}
(\P_{x}+\i\P_{y})\varphi (\bbox{x})|\F\rangle =-{\i\over \ell _{B}}{\partial \over \partial z^{*}}\varphi (\bbox{x})|\F\rangle  = 0 ,
\label{LLLcondiDress}
\end{equation}
where $\P_{j}$ is the covariant momentum for the new \CB field $\varphi (\bbox{x})$: 
See (\ref{CovarMomenR}).  We solve (\ref{EqForA}) as 
\begin{equation}
\a(\bbox{x})= m\int \dy \ln\biggl({|\bbox{x}-\bbox{y}|\over 2\ell _{B}}\biggr) \rho (\bbox{y}) -{|z|^{2}} .
\label{DressA}
\end{equation}
The $N$-body wave function $\F_{\varphi }[\bbox{x}]$ of dressed composite bosons is obtained by 
solving the LLL condition (\ref{LLLcondiDress}), and it is given simply by an 
analytic function $\omega [z]$ as in (\ref{WaveFunctDress}).  We shall verify in the next 
section that one analytic function $\omega [z]$ characterizes one state $|\F\rangle $ as in 
(\ref{WaveElect}) in terms of electrons, or as in (\ref{NaiveBoson}) in terms of bare 
composite bosons, or as in (\ref{WaveFunctDress}) in terms of dressed composite 
bosons.  

The effective magnetic potential for the bare composite boson is 
$A^{\text{ext}}_{k}+(1/e)\partial _{k}\Theta $, which is rewritten as $(1/e)\varepsilon _{kj}\partial _{j}\a(\bbox{x})$.  Bare 
composite bosons feel the effective magnetic field $\B_{\text{eff}}(\bbox{x})$,
\begin{equation}
\B_{\text{eff}}(\bbox{x}) = e^{-1}\bbox{\nabla }^{2}\a(\bbox{x}) = m\phi _{D}\rho (\bbox{x}) - B.
\label{EffecMagne}
\end{equation}
The effective field vanishes, $\langle \B_{\text{eff}}\rangle =0$, on the homogeneous state 
$\langle \rho (\bbox{x})\rangle =\rho _{0}$ if $m\phi _{D}\rho _{0}=B$.  It occurs at the filling factor $\nu  \equiv  \rho _{0}\phi _{D}/B = 1/m$.  
At $\nu =1/m$ we rewrite the effective magnetic field as
\begin{equation}
e\B_{\text{eff}}(\bbox{x}) = \bbox{\nabla }^{2}\a(\bbox{x}) = 2\pi m \varrho  (\bbox{x}) ,
\label{StepA}
\end{equation}
with $\varrho  (\bbox{y})\equiv \rho (\bbox{y})-\rho _{0}$.  It is solved as
\begin{equation}
\a(\bbox{x}) = m\int \dy \ln\biggl({|\bbox{x}-\bbox{y}|\over 2\ell _{B}}\biggr) \varrho  (\bbox{y}) .
\label{SpinB}
\end{equation}
This formula is equivalent to (\ref{DressA}) at $\nu =1/m$ up to an irrelevant 
integration constant.  Since the field $\varphi (\bbox{x})$ is constructed by dressing a 
cloud of the effective magnetic field, we have termed it the dressed \CB 
field.
\section{Quantum Hall States}\label{SecQHS}

We derive the Hamiltonian in terms of dressed composite bosons by 
substituting (\ref{DressField}) into (\ref{Hamil:B}),
\begin{equation}
H = {1\over 2M}\int \dx \varphi ^{\ddag }(\bbox{x})(\P_{x}-\i\P_{y})(\P_{x}+\i\P_{y})\varphi (\bbox{x}) + {N\over 2}\omega _{c} + H_{C} ,
\label{HamilCB}
\end{equation}
where we have defined
\begin{equation}
\varphi ^{\ddag }(\bbox{x}) \equiv  \varphi ^{\dagger }(\bbox{x})\e^{2\a(\bbox{x})} ,
\label{DressOperaB}
\end{equation}
with which $\rho (\bbox{x})=\psi ^{\dagger }(\bbox{x})\psi (\bbox{x})=\varphi ^{\ddag }(\bbox{x})\varphi (\bbox{x})$.  The covariant momentum 
$\P_{j}=P_{j}+\partial _{j}\Theta -\i\partial _{j}\a$ reads,
\begin{equation}
\P_{j} =  -\i\partial _{j} + e(\delta _{jk} - \i\varepsilon _{jk})\a_{k},\quad  \a_{k}(\bbox{x})=-{1\over e}\varepsilon _{kj}\partial _{j}\a(\bbox{x}).
\label{CovarMomenR}
\end{equation}
As we noticed in (\ref{LLLcondiDress}), it yields
\begin{equation}
\P_{x}+\i\P_{y}=-{\i\over \ell _{B}}{\partial \over \partial z^{*}} ,
\label{CoviaDress}
\end{equation}
with which the Hamiltonian (\ref{HamilCB}) is rewritten as
\begin{equation}
H = {\omega _{c}\over 2}\int \dx \biggl({\partial \over \partial z^{*}}\varphi (\bbox{x})\biggr)^{\dagger }\e^{2\a(\bbox{x})}{\partial \over \partial z^{*}}\varphi (\bbox{x}) + {N\over 2}\omega _{c} + H_{C} .
\label{HamilMono}
\end{equation}
It is manifestly hermitian.  

The Lagrangian density is
\begin{equation}
\L = \psi ^{\dagger }(\i\partial _{t}-eA^{\text{ext}}_{t})\psi  - \H = \varphi ^{\ddag }(\i\partial _{t}-eA^{\text{ext}}_{t}-\partial _{t}\Theta +\i\partial _{t}\a)\varphi  - \H ,
\end{equation}
where $\H$ is the Hamiltonian density and 
$A^{\text{ext}}_{\mu }=(A^{\text{ext}}_{t},A^{\text{ext}}_{k})$ is the potential of the external 
electromagnetic field.  Because the canonical conjugate of $\varphi (\bbox{x})$ is not 
$\i\varphi ^{\dagger }(\bbox{x})$ but $\i\varphi ^{\ddag }(\bbox{x})$, the equal-time canonical commutation relations are
\begin{equation}
[\varphi (\bbox{x}), \varphi ^{\ddag }(\bbox{y})] = \delta (\bbox{x}-\bbox{y}),\qquad [\varphi (\bbox{x}), \varphi (\bbox{y})] = [\varphi ^{\ddag }(\bbox{x}), \varphi ^{\ddag }(\bbox{y})] = 0.
\label{CCRbosonD}
\end{equation}
They are also derived \cite{RajaramanCB} by an explicit calculation from those of 
the electron fields $\psi (\bbox{x})$ and $\psi ^{\dagger }(\bbox{x})$ based on the definition (\ref{DressField}). 

The CB wave function is defined by
\begin{equation}
\F_{\varphi }[\bbox{x}]=\langle 0|\varphi (\bbox{x}_{1})\varphi (\bbox{x}_{2}) \cdots  \varphi (\bbox{x}_{N})|\F\rangle  .
\label{CompoStepADress}
\end{equation}
The LLL condition (\ref{LLLcondiDress}) implies that the wave function $\F_{\varphi }[\bbox{x}]$ is 
an analytic function, $\F_{\varphi }[\bbox{x}]=\omega [z]$.  With the use of the formula (\ref{DressA}) it 
is an easy exercise to derive the following relation \cite{RajaramanCB},
\begin{equation}
\varphi ^{\ddag }(\bbox{x}_{1})\varphi ^{\ddag }(\bbox{x}_{2})\cdots \varphi ^{\ddag }(\bbox{x}_{N})|0\rangle  = \Laugh \psi ^{\dagger }(\bbox{x}_{1})\psi ^{\dagger }(\bbox{x}_{2})\cdots \psi ^{\dagger }(\bbox{x}_{N})|0\rangle  ,
\label{DressB}
\end{equation}
where $\Laugh$ is the Laughlin function (\ref{LaughWave}).

Because of the commutation relations (\ref{CCRbosonD}) the state $|\F\rangle $ 
associated with the CB wave function (\ref{CompoStepADress}) is given by
\begin{eqnarray}
|\F\rangle &=&\int [\text{d}\bbox{x}] \F_{\varphi }[\bbox{x}] \varphi ^{\ddag }(\bbox{x}_{1})\varphi ^{\ddag }(\bbox{x}_{2}) \cdots  \varphi ^{\ddag }(\bbox{x}_{N})|0\rangle   \label{NbodyRb}\\
&=&\int [\text{d}\bbox{x}] \omega [z]\Laugh \psi ^{\dagger }(\bbox{x}_{1})\psi ^{\dagger }(\bbox{x}_{2}) \cdots  \psi ^{\dagger }(\bbox{x}_{N})|0\rangle  , \label{NbodyRc}
\end{eqnarray}
where use was made of (\ref{DressB}), and $[\text{d}\bbox{x}]=\dx_{1}\dx_{2}\cdots \dx_{N}$.  It follows 
from (\ref{NbodyRc}) that the electron wave function is $\F[\bbox{x}]=\omega [z]\Laugh$, as 
verifies the basic formulas (\ref{WaveElect}) $\sim $ (\ref{WaveFunctDress}).

The semiclassical ground state is the one that minimizes the total 
energy $\langle H\rangle $.  The Coulomb energy (\ref{CouloEnergPre}) is minimized by the state 
where $\langle \varrho  (\bbox{x})\rangle =0$.  It is realized when the electron density is homogeneous, 
\begin{equation}
\langle \rho (\bbox{x})\rangle =\langle \varphi ^{\ddag }(\bbox{x})\varphi (\bbox{x})\rangle =\e^{2\langle \a(\bbox{x})\rangle }\langle \varphi ^{\dagger }(\bbox{x})\varphi (\bbox{x})\rangle  =\rho _{0} .
\label{DressStepRho}
\end{equation}
In the semiclassical approximation we obtain
\begin{equation}
\prod _{r=1}^{N}\langle \rho (\bbox{x}_{r})\rangle  = \prod _{r=1}^{N}\e^{2\langle \a(\bbox{x}_r)\rangle }\langle \F|\varphi ^{\dagger }(\bbox{x}_{N})\cdots \varphi ^{\dagger }(\bbox{x}_{2})\varphi ^{\dagger }(\bbox{x}_{1})\varphi (\bbox{x}_{1})\varphi (\bbox{x}_{2})\cdots \varphi (\bbox{x}_{N})|\F\rangle .
\end{equation}
We insert a complete set $\sum |n\rangle \langle n|=1$ between two operators $\varphi ^{\dagger }(\bbox{x}_{1})$ and 
$\varphi (\bbox{x}_{1})$.  When the state $|\F\rangle $ contains $N$ electrons, only the vacuum term 
$|0\rangle \langle 0|$ survives in the complete set because $\varphi (\bbox{x}_{r})$ decreases the electron 
number by one.  Hence, $N$-body wave function (\ref{CompoStepADress}) is given by
\begin{equation}
\F_{\varphi }[\bbox{x}] = \rho _{0}^{N/2} \prod _{r=1}^{N}\e^{-\langle \a(\bbox{x}_r)\rangle } ,
\end{equation}
up to an irrelevant phase factor.  On the other hand, to suppress the kinetic 
energy we impose the LLL condition (\ref{LLLcondiDress}), as requires $\F_{\varphi }[\bbox{x}]$ to 
be analytic.  Consequently, it follows that $\langle \a(\bbox{x})\rangle =\text{constant}$ or 
$\B_{\text{eff}}=0$, which is possible only at $\nu =1/m$ from (\ref{EffecMagne}).  
Namely, the ground state is realized only at the magic filling factor $\nu =1/m$.  
At $\nu =1/m$ the ground-state wave function is given by $\F_{\varphi }[\bbox{x}]=\text{constant}$ 
in terms of \CBs, and therefore by the Laughlin wave function (\ref{LaughWave}) in 
terms of electrons.  In this way the Laughlin state is proved to be the ground 
state in the improved \CB theory.

For the sake of completeness, we briefly recall the results 
\cite{LGCSx} of the corresponding analysis with use of bare composite bosons.  We 
can derive the equations similar to (\ref{NbodyRb}) and (\ref{NbodyRc}), which verifies 
the mapping (\ref{NaiveBoson}).  The semiclassical ground state is similarly given 
by $\F_{\phi }(\bbox{x})=\text{constant}$ in terms of bare composite bosons.  The wave 
function in terms of electrons is singular,
\begin{equation}
\F[\bbox{x}]=\e^{\i m\sum _{r<s}\theta (z_{r}-z_{s})} . \label{NaiveGrounWave}
\end{equation}
The state does not belong to the \LLLd.  We can remove this singular 
short-distance behavior by including a higher order perturbation correction 
\cite{LGCSx,EIcoher}.  Nevertheless, it makes the naive CB theory less attractive.  
\section{Semiclassical Analysis}\label{SecTE}

We analyze excitations on the QH state.  A priori two types of 
excitations are possible, that is, perturbative and nonperturbative ones in 
terms of the density fluctuation $\varrho  (\bbox{x})$ and its conjugate phase $\chi (\bbox{x})$.  To 
carry out a perturbative analysis, we parametrize the bare field as $\phi (\bbox{x}) = 
\e^{\i\chi (\bbox{x})} \sqrt {\rho _{0}+\varrho  (\bbox{x})}$ in terms of the density deviation $\varrho  (\bbox{x})$ and its 
canonical phase $\chi (\bbox{x})$.  The dressed field $\varphi (\bbox{x})$ is a nonlocal operator due 
to the factor $\e^{-\a(\bbox{x})}$ with (\ref{SpinB}),
\begin{equation}
\varphi (\bbox{x}) = \e^{-\a(\bbox{x})}\e^{\i\chi (\bbox{x})} \sqrt {\rho _{0}+\varrho  (\bbox{x})} .
\label{BareDress}
\end{equation}
Substituting (\ref{BareDress}) into the Hamiltonian (\ref{HamilCB}) and expanding 
various quantities in term of $\varrho  (\bbox{x})$ and $\chi (\bbox{x})$, we obtain the perturbative 
Hamiltonian, which is found to be identical to the one analyzed already in the 
bare \CB theory \cite{LGCSx}, as should be the case.  The result is that there 
exist no perturbative fluctuations confined to the \LLLd.  We conclude that 
all excitations in the \LLL are nonperturbative objects.  The improved theory 
confirms this assertion by showing explicitly how they are created on the 
ground state.  Indeed, any excited state is represented as in (\ref{NbodyRb}), 
which is a nonlocal object because the creation operator $\varphi ^{\ddag }(\bbox{x})$ is a nonlocal 
operator as in (\ref{BareDress}).
\subsection{Vortices}

We first examine excited states when the wave function 
(\ref{CompoStepADress}) is factorizable, $\F_{\varphi }[\bbox{x}]=\omega [z]=\prod _{r}\omega (z_{r})$.  In this case we 
can easily make a semiclassical analysis of the one-point function 
$\langle \varphi (\bbox{x})\rangle =\omega (z)$ by setting
\begin{equation}
\e^{-\a(\bbox{x})}\e^{\i\chi (\bbox{x})} \sqrt {\rho _{0}+\varrho  (\bbox{x})} = \omega (z) ,
\label{ClassGenerA}
\end{equation}
based on the parametrization (\ref{BareDress}).  Here and hereafter, we use the 
same symbols $\a(\bbox{x})$, $\varrho  (\bbox{x})$ and $\chi (\bbox{x})$ also for the classical fields.  When 
an analytic function $\omega (z)$ is given, (\ref{ClassGenerA}) is an integral equation 
determining the density deviation $\varrho  (\bbox{x})$ confined to the \LLLd. 

We transform (\ref{ClassGenerA}) into a differential equation.  The 
Cauchy-Riemann equation for the analytic function (\ref{ClassGenerA}) yields,
\begin{equation}
\partial _{j}\bigl(\a(\bbox{x}) - \ln \sqrt {\rho _{0}+\varrho  (\bbox{x})}\bigr) = -\varepsilon _{jk}\partial _{k}\chi (\bbox{x}) .
\label{StepB}
\end{equation}
Using (\ref{StepA}) we obtain that
\begin{equation}
{\nu \over 4\pi }\bbox{\nabla }^{2}\ln\biggl(1+{\varrho  (\bbox{x})\over \rho _{0}}\biggr) - \varrho  (\bbox{x}) = \nu Q^{V}(\bbox{x}) ,
\label{SolitEqVorte}
\end{equation}
which we call the soliton equation, where
\begin{equation}
Q^{V}(\bbox{x}) = {1\over 2\pi }\varepsilon _{jk}\partial _{j}\partial _{k}\chi (\bbox{x}) = {1\over 2\pi \i}\varepsilon _{jk}\partial _{j}\partial _{k}\ln \omega (z) 
\label{VorteCharg}
\end{equation}
is the topological charge density associated with the excitation.  This is 
nonvanishing since $\ln\omega (z)$ is a multivalued function unless 
$\omega (z)=\text{constant}$.  The topological current is
\begin{equation}
Q_{\mu }^{V}(\bbox{x}) = {1\over 2\pi }\varepsilon _{\mu \nu \lambda }\partial _{\nu }\partial _{\lambda }\chi (\bbox{x}) ,
\end{equation}
with $Q^{V}(\bbox{x})=Q^{V}_{0}(\bbox{x})$.  It is a conserved quantity, $\partial ^{\mu }Q_{\mu }^{V}(\bbox{x})=0$.  Topological 
solitons are generated around the zeros of $\omega (z)$ according to the soliton 
equation (\ref{SolitEqVorte}).  The density modulation is induced in order to 
confine the excitation within the \LLLd.

The topological charge is evaluated as
\begin{equation}
Q^{V} = \int \dx Q^{V}(\bbox{x}) = {1\over 2\pi \i}\oint \text{d}x_{k} \partial _{k}\ln \omega (z) ,
\label{VorteChargA}
\end{equation}
where the loop integration $\oint $ is made to encircle the excitation at infinity 
($|\bbox{x}|\rightarrow \infty $) provided $\omega (z)$ is regular everywhere.  The topological charge is 
uniquely determined by the asymptotic behavior of the classical field $\varphi (\bbox{x})$,
\begin{equation}
\varphi (\bbox{x}) \rightarrow  \sqrt {\rho _{0}}z^{q} ,\quad \quad \text{as}\quad  |z| \rightarrow  \infty  ,
\label{VorteAsymp}
\end{equation}
for which the electron number is
\begin{equation}
\Delta N = \int \dx \varrho  (\bbox{x}) = -\nu  Q_{\text{V}} = -\nu q ,
\label{ElectNumbe}
\end{equation}
as follows from the soliton equation (\ref{SolitEqVorte}).  The electric charge 
carried by this soliton is $-e\Delta N=\nu qe$.  It is a hole made in the condensate of 
composite bosons.  We may as well derive the electric number (\ref{ElectNumbe}) 
more directly from the parametrization (\ref{ClassGenerA}).  We take the asymptotic 
behavior $|z|\rightarrow \infty $ in (\ref{ClassGenerA}) and equate it with (\ref{VorteAsymp}),
\begin{equation}
\varrho  (\bbox{x})\rightarrow 0,\quad \quad  \chi (\bbox{x})\rightarrow q\theta , \quad \quad  \a(\bbox{x})\rightarrow -q\ln|z|, \quad \quad \text{as}\quad  |z| \rightarrow  \infty  .
\label{AsympChara}
\end{equation}
Taking the limit $|\bbox{x}|\rightarrow \infty $ in (\ref{SpinB}) we recover the quantization of the 
electron number (\ref{ElectNumbe}).  Actually we have determined the coefficient 
$\sqrt {\rho _{0}}$ in the asymptotic behavior (\ref{VorteAsymp}) in this way.

A topological excitation carries a quantized topological charge, which 
has to be created all at once.  It cannot be created by an accumulation of 
perturbative fluctuations, as agrees with the perturbative result mentioned at 
the begining of this section.  In the absence of the Coulomb term all these 
excitations are degenerate with the ground state, which explains the 
degeneracy in the lowest Landau level at $\nu <1$.  The degeneracy is removed 
since any density modulation acquires a Coulomb energy,
\begin{equation}
\langle H_{C}\rangle  = {e^{2}\over 2\varepsilon }\int \dx\dy {\varrho  (\bbox{x})\varrho  (\bbox{y})\over |\bbox{x}-\bbox{y}|} = \gamma \nu ^{2}q^{2}{e^{2}\over \varepsilon \ell _{B}} ,
\label{CouloEnerg}
\end{equation}
where $\gamma $ is a constant of order one.

\begin{figure}[hbt]
\epsfxsize=100mm\epsfbox{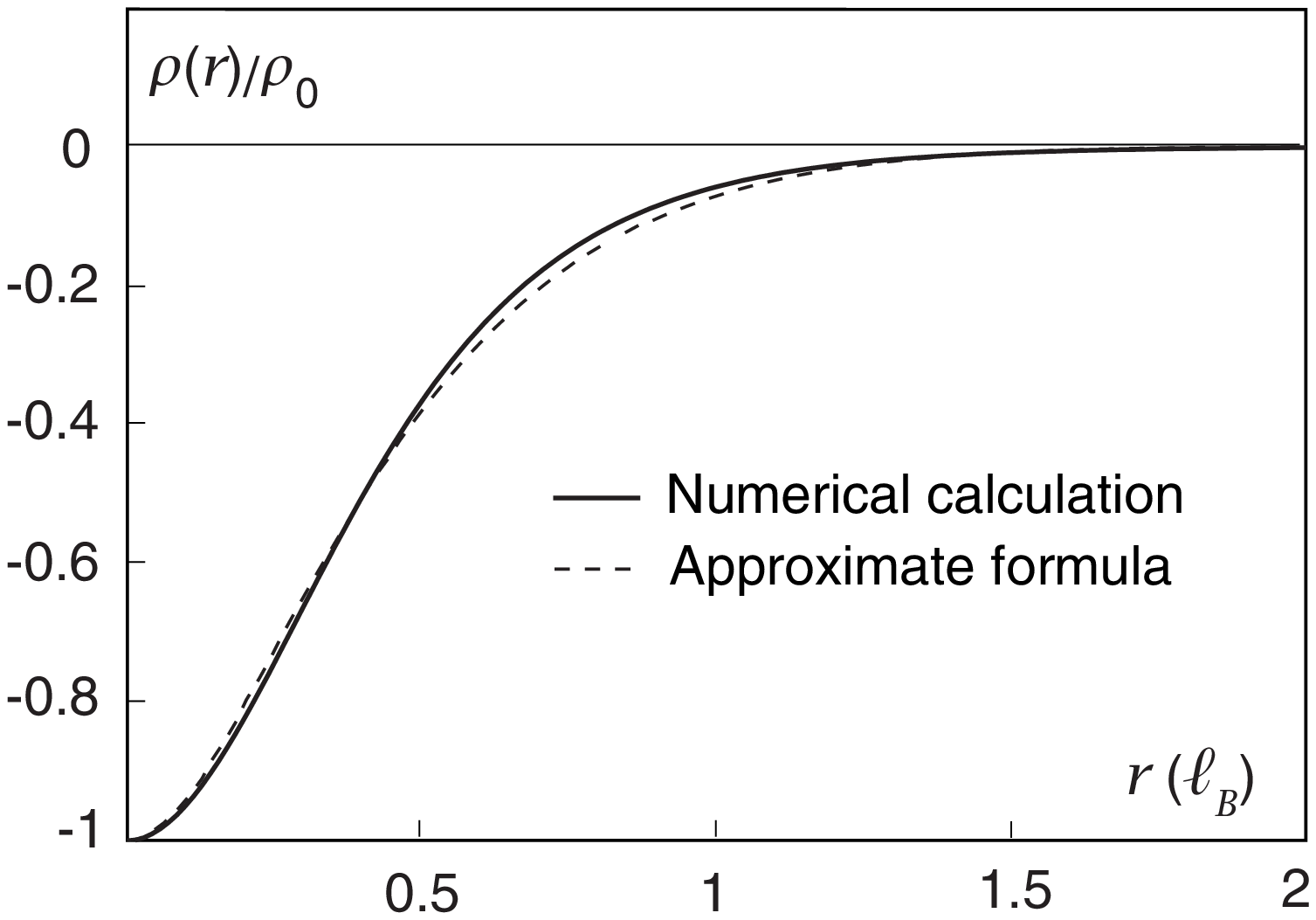}
\renewcommand{\baselinestretch}{0.8}
\caption{\small
The density moduration around a vortex with $q=1$ is plotted.  The solid curve 
is obtained by solving the differential equation (\ref{ModifLiuvi})  
numerically.
The dashed curve is drawn by using the approximate formula (\ref{BetteAppro}).  
}\label{VortNumePS}
\end{figure}

An explicit example is given by a vortex (quasihole) sitting at $\bbox{x}=0$, 
whose wave function is $\F_{\varphi }[\bbox{x}]=\prod _{r}z_{r}$ up to a normalization factor, or 
$\omega (z)=\sqrt {\rho _{0}}z$.  In this example the topological charge (\ref{VorteCharg}) is 
concentrated at the vortex center, $Q^{V}_{0}(\bbox{x})=\delta (\bbox{x})$.  A crude approximation of 
the soliton equation (\ref{SolitEqVorte}) reads
\begin{equation}
{\ell _{B}^{2}\over 2}\bbox{\nabla }^{2}\varrho  (\bbox{x}) - \varrho  (\bbox{x}) = \nu \delta (\bbox{x}) ,
\end{equation}
since $\nu =2\pi \rho _{0}\ell _{B}^{2}$.  Its exact solution \cite{LaughlinA} is $\varrho  (\bbox{x})=-(\nu /\pi \ell _{B}^{2})K_{0}(s)$ 
with $s=\sqrt {2}r/\ell _{B}$ and $K_{0}(s)$ the modified Bessel function.  This is a rather 
poor approximation because of its singular behavior, $\varrho  (\bbox{x})\rightarrow -\infty $, at the vortex 
center ($\bbox{x}=0$).  A better approximation is given by
\begin{equation}
\varrho  (\bbox{x}) = -\rho _{0}\biggl(1+s-{s^{2}\over 6}\biggr)\e^{-s} ,
\label{BetteAppro}
\end{equation}
which has the correct behavior both at $\bbox{x}=0$ and $|\bbox{x}|\gg \ell _{B}$.  Furthermore, it 
has the correct topological charge.  For a numerical analysis, it is 
convenient to set $\rho (\bbox{x})=\rho _{0}\e^{u(s)}$ in the soliton equation (\ref{SolitEqVorte}), as 
yields \cite{EHIc},
\begin{equation}
{d^{2}u\over ds^{2}} + {1\over s}{du\over ds} + 1 = \e^{u(s)} ,
\label{ModifLiuvi}
\end{equation}
for $s>0$.  The result of a numerical analysis shows that the density 
modulation is well approximated by (\ref{BetteAppro}), as in Fig.\ref{VortNumePS}.  
The Coulomb energy is given by (\ref{CouloEnerg}) with $\gamma \simeq 0.39$.

\section{Quantum Hall Ferromagnet}\label{SecQHFerro}

We proceed to analyze the QH system with the SU(2) symmetry.  The 
electron field $\psi ^{\alpha }(\bbox{x})$ has the index $\alpha =\upA ,\dnA $.  It denotes the electron spin in 
the monolayer QH system with the spin SU(2) symmetry, or the layer index in a 
certain bilayer QH system with the pseudospin SU(2) symmetry.  For 
definiteness we analyze the monolayer spin system in this paper.
\subsection{Hamiltonian}

The Hamiltonian depends on the electron spin through the Zeeman energy 
term,
\begin{equation}
H_{Z} = -g^{*}\mu _{B}B \int \dx S^{z}(\bbox{x}) ,
\label{ZeemaTerm}
\end{equation}
with $S^{z}={1\over 2}(\psi ^{\upA \dagger }\psi ^{\upA }-\psi ^{\dnA \dagger }\psi ^{\dnA })$, where $g^{*}$ is the gyromagnetic factor and $\mu _{B}$ 
the Bohr magneton.  Each Landau level contains two energy levels with the 
one-particle gap energy $g^{*}\mu _{B}B$.  The lowest Landau level is filled at $\nu =2$.  
We consider the case where the Zeeman energy is much smaller than the Coulomb 
energy.  Though one Landau level contains two degenerate energy levels in the 
vanishing limit of the Zeeman coupling ($g^{*}=0$), the system becomes 
incompressible at $\nu =1/m$.  The physical reason is the Coulomb exchange 
energy, as we now see.

We define the \textit{bare} CB field $\phi ^{\alpha }(\bbox{x})$ and the \textit{dressed} 
CB field $\varphi ^{\alpha }(\bbox{x})$ by 
\begin{equation}
\phi ^{\alpha }(\bbox{x}) = \e^{-\i\Theta (\bbox{x})}\psi ^{\alpha }(\bbox{x}),\quad  \varphi ^{\alpha }(\bbox{x}) = \e^{-\a(\bbox{x})}\phi ^{\alpha }(\bbox{x}),
\end{equation}
where the phase field $\Theta (\bbox{x})$ and the auxiliary field $\a(\bbox{x})$ are given by 
(\ref{PhaseTheta}) and (\ref{SpinB}), respectively, with the total electron density 
$\rho (\bbox{x})=\sum _{\alpha }\psi ^{\alpha \dagger }(\bbox{x})\psi ^{\alpha }(\bbox{x})=\sum _{\alpha }\varphi ^{\alpha \dagger }(\bbox{x})e^{2\a(\bbox{x})}\varphi ^{\alpha }(\bbox{x})$.  The Hamiltonian is
\begin{eqnarray}
H &=&{1\over 2M}\sum _{\alpha }\int \dx \psi ^{\alpha \dagger }(\bbox{x})(P_{x}^{2}+P_{y}^{2})\psi ^{\alpha }(\bbox{x}) + H_{C}+H_{Z}  \label{HamilSpinA}\\
&=&\omega _{c }\sum _{\alpha }\int \dx \biggl({\partial \over \partial z^{*}}\varphi ^{\alpha }(\bbox{x})\biggr)^{\dagger }e^{2\a(\bbox{x})}{\partial \over \partial z^{*}}\varphi ^{\alpha }(\bbox{x}) +{N\over 2}\omega _{c}+H_{C}+H_{Z} ,
\label{HamilSpinB}
\end{eqnarray}
with the Coulomb term $H_{C}$ and the Zeeman term $H_{Z}$.  The Coulomb term depends 
on the deviation $\varrho  (\bbox{x})$ of the total electron density from the average 
density, $\varrho  (\bbox{x})=\rho (\bbox{x})-\rho _{0}$, as in (\ref{CouloEnergPre}).
\subsection{Wave Function}

We may decompose the bare \CB field into the U(1) field $\phi (\bbox{x})$ and the 
SU(2) field $n^{\alpha }(\bbox{x})$, 
\begin{equation}
\phi ^{\alpha }(\bbox{x}) = \phi (\bbox{x})n^{\alpha }(\bbox{x}), \quad \sum _{\alpha }n^{\alpha \dagger }(\bbox{x})n^{\alpha }(\bbox{x})= 1 .
\label{BareSpin}
\end{equation}
The field $n^{\alpha }(\bbox{x})$ is the complex-projective (CP) field\cite{Skyrmion}, whose 
overall phase has been removed and given to the U(1) field $\phi (\bbox{x})$.  The spin 
operator is expressed as 
\begin{equation}
S^{a}(\bbox{x})={1\over 2}\rho (\bbox{x})\Sigma ^{a}(\bbox{x}) ,
\end{equation}
where
\begin{equation}
\Sigma ^{a}(\bbox{x})=\bbox{n}^{\dagger }(\bbox{x}){\tau ^{a}}\bbox{n}(\bbox{x}), \quad  \bbox{n}(\bbox{x}) = \bpmatrix n^{\upA }(\bbox{x})\\n^{\dnA }(\bbox{x})\epmatrix.
\end{equation}
In terms of the dressed CB field the decomposition reads
\begin{equation}
\varphi ^{\alpha }(\bbox{x}) = \varphi (\bbox{x})n^{\alpha }(\bbox{x}), \quad  \varphi (\bbox{x}) = \e^{-\a(\bbox{x})}\phi (\bbox{x}) .
\label{SpinAD}
\end{equation}
The SU(2) component $n^{\alpha }(\bbox{x})$ is common between the bare and dressed fields 
(\ref{BareSpin}) and (\ref{SpinAD}):  It is a local field.  On the other hand, $\varphi (\bbox{x})$ is 
a nonlocal field due to the factor $\e^{-\a(\bbox{x})}$ as in the spinless theory.

The ground state minimizes both the Coulomb and Zeeman energies.  The 
Coulomb energy is minimized by the homogeneous electron density, $\langle \rho (\bbox{x})\rangle =\rho _{0}$.  
The Zeeman energy is minimized when all electrons are polarized into the 
positive $z$ axis, $\langle n^{\upA }(\bbox{x})\rangle =1$ and $\langle n^{\dnA }(\bbox{x})\rangle =0$.  The ground state is unique, 
which we denote by $|g_{0}\rangle $.

The two-component CB field is $\Phi (\bbox{x})=\varphi (\bbox{x})\bbox{n}(\bbox{x})$.  With the Hamiltonian 
(\ref{HamilSpinB}), the LLL condition for the state $|\F\rangle $ is
\begin{equation}
{\partial \over \partial z^{*}}\Phi (\bbox{x})|\F\rangle  = 0 .
\label{LLLcondiBL}
\end{equation}
Because of this condition the $N$-body wave function is analytic,
\begin{equation}
\F_{\varphi }[\bbox{x}] = \langle 0|\Phi (\bbox{x}_{1})\Phi (\bbox{x}_{2}) \cdots  \Phi (\bbox{x}_{N})|\F\rangle  = \Omega [z] ,
\label{DressWaveFunct}
\end{equation}
where $\Omega [z]$ is totally symmetric in $N$ variables.  When it is factorizable, 
$\Omega [z]=\prod _{r}\Omega (z_{r})$, the electron wave function reads
\begin{equation}
\F[\bbox{x}] = \prod _{r}\bpmatrix \omega ^{\upA }(z_{r})\\ \omega ^{\dnA }(z_{r})\epmatrix \Laugh.
\end{equation}
The ground-state wave function is
\begin{equation}
\F[\bbox{x}] = \prod _{r}\bpmatrix 1\\ 0\epmatrix \Laugh,
\label{VacuuMono}
\end{equation}
where all spins are polarized.
\subsection{Spin Texture}

We consider a spin texture given by performing an SU(2) transformation 
on the ground state $|g_{0}\rangle $,
\begin{equation}
|\F\rangle  = \e^{\i\O}|g_{0}\rangle  ,
\label{SpinTextu}
\end{equation}
where $\O$ is its generator,
\begin{equation}
\O = \sum _{a}\int \dx f^{a}(\bbox{x}) S^{a}(\bbox{x}) = \sum _{a}\int {\dq}f^{a}_{-\bbox{q}}S^{a}_{\bbox{q}}  .
\label{GenerSU}
\end{equation}
The spin texture is described by the classical sigma field 
$s^{a}(\bbox{x})=\langle \F|\Sigma ^{a}(\bbox{x})|\F\rangle $, which we parametrize as
\begin{eqnarray}
s^{x}(\bbox{x})=\sigma (\bbox{x}),  \quad  s^{y}(\bbox{x})= \sqrt {1-\sigma ^{2}(\bbox{x})}\sin\vartheta (\bbox{x}), \quad  s^{z}(\bbox{x})=\sqrt {1-\sigma ^{2}(\bbox{x})}\cos\vartheta (\bbox{x}). 
\label{ClassPS}
\end{eqnarray}
It is classified by the Pontryagin number \cite{Skyrmion}, $Q^{P}=\int \dx Q^{P}_{0}(\bbox{x})$, with 
the topological current
\begin{equation}
Q_{\mu }^{P}(\bbox{x}) = {1\over 8\pi }\varepsilon _{abc}\varepsilon _{\mu \nu \lambda }s_{a}\partial ^{\nu }s_{b}\partial ^{\lambda }s_{c}  .
\end{equation}
It is absolutely conserved, $\partial ^{\mu }Q_{\mu }^{P}=0$.  

The spin texture (\ref{SpinTextu}) does not belong to the \LLLd.  The 
excitation energy is naively given by $\langle \Psi |H|\Psi \rangle $ with the Hamiltonian 
(\ref{HamilSpinB}), to which the kinetic term and the Zeeman term contribute but 
the Coulomb term does not.  The kinetic energy is of the order of $\omega _{c}$, as 
implies the spin stiffness of this order.  Such an excitation is impossible at 
sufficiently low temperature.  It is necessary to excite merely the component 
$|\FF\rangle $ of $|\F\rangle $ belonging to the \LLLd.  Furthermore, it is also necessary 
to extract the LLL component of the potential term since it kicks out the 
state out of the \LLLd.  
\subsection{LLL Projection}

The LLL component $|\FF\rangle $ is extracted from the spin texture $|\F\rangle $ by 
extracting the LLL component $\llangle f^{a}(\bbox{x})\rrangle $ from the ``wave packet'' $f^{a}(\bbox{x})$ in the 
generator $\O$ of the SU(2) transformation (\ref{GenerSU}).  This turns out to 
replace the plane wave $\e^{\i{\bbox{x}\bbox{q}}}$ with \cite{refLLL,EzaIQC}
\begin{equation}
\llangle \e^{\i{\bbox{x}\bbox{q}}}\rrangle \equiv \e^{-{1\over 4}\bbox{q}^{2}\ell _{B}^{2}}\e^{\i{\bbox{X}\bbox{q}}} 
\end{equation}
in the Fourier representation of $f^{a}(\bbox{x})$, where $\bbox{X}=(X,Y)$ is the guiding 
center.  We call $\llangle \e^{\i{\bbox{x}\bbox{q}}}\rrangle $ the LLL projection of $\e^{\i{\bbox{x}\bbox{q}}}$.  The generator 
(\ref{GenerSU}) is projected as
\begin{equation}
\OO = \sum _{a}\int \dx \llangle f^{a}(\bbox{x})\rrangle  S^{a}(\bbox{x}) = \sum _{a}\int {\dq}f^{a}_{-\bbox{q}}\widehat{S}^{a}_{\bbox{q}}  ,
\end{equation}
where
\begin{equation}
\widehat{S}^{a}_{\bbox{q}}=(2\pi )^{-1}\int \dx \llangle \e^{-\i{\bbox{q}\bbox{x}}}\rrangle S^{a}(\bbox{x}).
\end{equation}
Similarly we define
\begin{equation}
\widehat{\rho }_{\bbox{q}}=(2\pi )^{-1}\int \dx \llangle \e^{-\i{\bbox{q}\bbox{x}}}\rrangle \rho (\bbox{x}) 
\end{equation}
for the electron density operator.  

From the commutation relation $[X,Y]=-i\ell _{B}^{2}$ between the $X$ and $Y$ 
components of the guiding center, we obtain
\begin{equation}
[\e^{\i\bbox{q}\bbox{X}}, \e^{\i\bbox{p}\bbox{X}}] = 2\i\e^{\i(\bbox{q}+\bbox{p})\bbox{X}} \sin\bigl[\ell _{B}^{2}{\bbox{q}\!\wedge\!\bbox{p}\over 2}\bigr], \label{Walgebra}
\end{equation}
with $\bbox{q}\!\wedge\!\bbox{p}\equiv q_{x}p_{y}-q_{y}p_{x}$.  The translation $\e^{\i\bbox{q}\bbox{x}}$ is Abelian, but the magnetic 
translation $\e^{\i\bbox{q}\bbox{X}}$ is non-Abelian.  It governs the symmetric structure of 
the two-dimensional space after the LLL projection.  It is straightforward to 
derive the following W$_{\infty }\times $SU(2) algebra \cite{EzaIQC},
\begin{mathletters}\label{SUCommu}
\begin{eqnarray}
&[\widehat{\rho }_{\bbox{p}},\widehat{\rho }_{\bbox{q}}]={\i\over \pi }\widehat{\rho }_{\bbox{p}+\bbox{q}}\sin\bigl[\Wed{\bbox{p}}{\bbox{q}}\bigr]\exp\bigl[{\ell _{B}^{2}\over 2}\bbox{p}\bbox{q}\bigr], \label{SUCommuA}\\
&[\widehat{S}^{a}_{\bbox{p}},\widehat{\rho }_{\bbox{q}}]={\i\over \pi }\widehat{S}^{a}_{\bbox{p}+\bbox{q}}\sin\bigl[\Wed{\bbox{p}}{\bbox{q}}\bigr]\exp\bigl[{\ell _{B}^{2}\over 2}\bbox{p}\bbox{q}\bigr],\label{SUCommuB}\\
&[\widehat{S}^{a}_{\bbox{p}}, \widehat{S}^{b}_{\bbox{q}}]={\i\over 2\pi }\varepsilon ^{abc}\widehat{S}^{c}_{\bbox{p}+\bbox{q}}\cos\bigl[\Wed{\bbox{p}}{\bbox{q}}\bigr]\exp\bigl[{\ell _{B}^{2}\over 2}\bbox{p}\bbox{q}\bigr] \nonumber\\
&\hspace*{20mm}+{\i\over 4\pi }\delta ^{ab}\widehat{\rho }_{\bbox{p}+\bbox{q}}\sin\bigl[\Wed{\bbox{p}}{\bbox{q}}\bigr]\exp\bigl[{\ell _{B}^{2}\over 2}\bbox{p}\bbox{q}\bigr],
\label{SUCommuC}
\end{eqnarray}
\end{mathletters}
based on the algebra (\ref{Walgebra}) of the magnetic translation.

We make the LLL projection of the Hamiltonian (\ref{HamilSpinB}), whose 
result we denote by $\widehat{H}$.  The Coulomb term reads
\begin{equation}
\widehat{H}_{C} = {1\over 2}\int \dx\dy \varrho  (\bbox{x})\llangle V(\bbox{x}-\bbox{y})\rrangle \varrho  (\bbox{y}) = \pi  \int {\dq}V(\bbox{q}) \widehat{\rho }_{-\bbox{q}}\widehat{\rho }_{\bbox{q}} ,
\label{ProjeCoulo}
\end{equation}
where $V(\bbox{q})$ is the Fourier transformation of the potential $V(\bbox{x})=e^{2}/\varepsilon |\bbox{x}|$.
\subsection{Exchange Energy}

We evaluate the energy $\langle \FF|\widehat{H}|\FF\rangle $ by making a perturbative expansion 
of the spin texture around the ground state,
\begin{equation}
H_{\text{eff}}\equiv \langle \FF|\widehat{H}|\FF\rangle  = \langle g_{0}|\widehat{H}|g_{0}\rangle  - \langle g_{0}|[\OO, \widehat{H}]|g_{0}\rangle  + \cdots  .
\end{equation}
Making a straightforward algebraic calculation, making a gradient expansion 
and taking the lowest order term in $f^{a}_{-\bbox{q}}$, we obtain the exchange energy 
\cite{EzaIQC,Moon,KallinHalperin} from the Coulomb term (\ref{ProjeCoulo}),
\begin{equation}
H_{\text{stiff}}= {1\over 2}\rho _{s}\sum _{a}\int \dx [\partial _{k}s^{a}(\bbox{x})]^{2},
\label{SpinStiffTerm}
\end{equation}
where $s^{a}(\bbox{x})$ is the nonlinear sigma field with $\rho _{s}=\nu e^{2}/(16\sqrt {2\pi }\varepsilon \ell _{B})$.  It 
describes the spin stiffness.  Combining the exchange energy and the Zeeman 
energy we obtain the effective Hamiltonian,
\begin{equation}
H_{\text{eff}}= {1\over 2}\rho _{s}\sum _{a}\int \dx [\partial _{k}s^{a}(\bbox{x})]^{2} - {\rho _{0}\over 2}g^{*}\mu _{B}B \int \dx s^{z}(\bbox{x}) ,
\label{EnergChangSPN}
\end{equation}
The exchange energy arises because of the following reason:  The local spin 
rotation has components in higher Landau level since it is not a symmetry of 
the Hamiltonian.  Only its LLL component is excited at sufficiently low 
temperature.  Since the LLL components of the spin operators and the density 
operator do not commute as in (\ref{SUCommuB}), the local spin rotation induces a 
local density modulation and affects the Coulomb energy.  

The spin-stiffness term is precisely the nonlinear-sigma-model 
Hamiltonian \cite{SkyrmQH}.  Though the term is derived perturbatively in the 
present framework, there is much numerical evidence \cite{SkyrmQH,SkyrmA,RezayiA} 
that it captures correctly the long-distance physics associated with Skyrmion 
excitations.
\subsection{Goldstone Mode}

The Zeeman effect is quite small in actual samples.  We consider the 
vanishing limit of the Zeeman term ($g^{*}=0$).  According to the effective 
Hamiltonian (\ref{EnergChangSPN}), the energy is minimized for any constant value 
of the sigma field, $\bbox{s}(\bbox{x})=\bbox{s}_{0}=$constant.  Hence, there exists a degeneracy in 
the ground states as indexed by $\bbox{s}_{0}$.  The choice of a ground state implies a 
spontaneous magnetization, or a \textit{quantum Hall ferromagnetism}.  When a 
continuous symmetry is spontaneously broken, there should arise a gapless mode 
known as the Goldstone mode and quantum coherence develops spontaneously.  

When we include a small Zeeman effect, the ground state $|g_{0}\rangle $ is 
chosen where $\bbox{s}_{0}=(0,0,1)$.  We can treat the Zeeman interaction as a 
perturbation because it is much less important than the Coulomb interaction.  
The key property of the QH ferromagnet is that it is a macroscopic coherent 
state though the coherent length is finite, where all spin components are 
simultaneously measurable, $s^{a}(\bbox{x})=2\rho _{0}^{-1}\langle S^{a}(\bbox{x})\rangle $, with extremely good accuracy.  
An evidence is the existence of coherent excitations such as Skyrmions.

The Goldstone mode describes small fluctuations of the CP field around 
the ground state (\ref{VacuuMono}).  Up to the lowest order of the perturbation in 
the CP field $\bbox{n}(\bbox{x})$, it is parametrized as \cite{EzaIQC}
\begin{equation}
n^{\upA }(\bbox{x})=1, \quad \quad  n^{\dnA }(\bbox{x})={\zeta (\bbox{x})\over \sqrt {\rho _{0}}},
\label{CPGolds}
\end{equation}
with $[\zeta (\bbox{x}),\zeta ^{\dagger }(\bbox{y})]=\delta (\bbox{x}-\bbox{y})$.  The LLL condition (\ref{LLLcondiBL}) yields two 
conditions,
\begin{equation}
{\partial \over \partial z^{*}} \varphi (\bbox{x})|\F\rangle  = 0 ,\quad  {\partial \over \partial z^{*}} \zeta (\bbox{x})|\F\rangle  = 0 ,
\end{equation}
up to this order.  Although they look similar, they describe very different 
excitation modes.  As in the spinless QH system, $\varphi (\bbox{x})$ is a nonlocal field 
and generates extended objects.  On the other hand, $\zeta (\bbox{x})$ is a local field, 
and it describes the Goldstone mode.

We may relate $\zeta (\bbox{x})$ to the classical sigma field (\ref{ClassPS}),
\begin{equation}
\langle \zeta (\bbox{x})\rangle  = {\sqrt {\rho _{0}}\over 2}\bigl\{\sigma (\bbox{x})+\i\vartheta (\bbox{x})\bigr\} .
\label{GoldsPertu}
\end{equation}
The effective Hamiltonian (\ref{EnergChangSPN}) is recognized as a classical 
counterpart of the quantum version,
\begin{equation}
H_{\text{eff}}=  {2\rho _{s}\over \rho _{0}}\int \dx \bigl[\partial _{k}\zeta ^{\dagger }(\bbox{x})\partial _{k}\zeta (\bbox{x}) + \xi _{L}^{-2}\zeta ^{\dagger }(\bbox{x})\zeta (\bbox{x})\bigr],
\end{equation}
on the coherent state.  Here, $\xi _{L}$ is the coherent length,
\begin{equation}
\xi _{L}=\sqrt {2\rho _{s}\over g^{*}\mu _{B}B\rho _{0}} = {(2\pi )^{1/4}\ell _{B}\over 2\sqrt {2\widetilde{g}^{ }} },
\label{CoherLengt}
\end{equation}
where
\begin{equation}
\widetilde{g}={g^{*}\mu _{B}B\over e^{2}/\varepsilon \ell _{B}} 
\label{NormaGfacto}
\end{equation}
is the ratio of the Zeeman energy to the Coulomb energy.  We call it the 
normalized g-factor.  We have $\xi _{L}\sim 4\ell _{B}$ in typical samples at $B\simeq 10$ Tesla, 
where $\widetilde{g}\simeq 0.02$.  In the momentum space the effective Hamiltonian reads
\begin{equation}
\H_{\rm{eff}}(\bbox{k}) = E_{\bbox{k}}\zeta ^{\dagger }_{\bbox{k}}\zeta _{\bbox{k}} ,
\end{equation}
with $[\zeta _{\bbox{k}}, \zeta ^{\dagger }_{\bbox{l}}]=\delta (\bbox{k}-\bbox{l})$, and the dispersion relation is
\begin{equation}
E_{\bbox{k}} = {2\rho _{s}\over \rho _{0}}\bbox{k}^{2} + g^{*}\mu _{B}B.
\label{SuperModeZeema}
\end{equation}
The Goldstone mode has acquired a gap $E_{0}=g^{*}\mu _{B}B$.
\section{Topological Excitations}\label{SecSpinTE}

We analyze topological excitations on the QH ferromagnet.  We use the 
semiclassical approximation.  When the $N$-body wave function is factorizable, 
the one-point function is analytic, $\langle \varphi ^{\alpha }(\bbox{x})\rangle =\omega ^{\alpha }(z)$.  From (\ref{SpinAD}), the 
one-point function is parametrized as
\begin{equation}
\e^{-\a(\bbox{x})}\e^{\i\chi (\bbox{x})} \sqrt {\rho _{0}+\varrho  (\bbox{x})}n^{\alpha }(\bbox{x}) = \omega ^{\alpha }(z) ,
\label{ClassGenerB}
\end{equation}
since $|\phi (\bbox{x})|^{2}=\rho _{0}+\varrho  (\bbox{x})$.  Here and hereafter, all fields are classical fields.  
When the wave function $\omega ^{\alpha }(z)$ is given, the electron density $\varrho  (\bbox{x})$ and the 
spin field $S^{a}(\bbox{x})$ are determined by this equation.  There are two types of 
excitations associated with the U(1) part and the SU(2) part of the \CB field.  
The U(1) excitation has a characteristic length $\ell _{B}$, while the SU(2) 
excitation has no scale provided the Zeeman term is neglected.
\subsection{Vortex Excitations}

The U(1) excitation is generated on the spin-polarized ground state 
(\ref{VacuuMono}) when $\partial _{k}\chi (\bbox{x})\not=0$ and $\partial _{k}n^{\alpha }(\bbox{x})=0$ in (\ref{ClassGenerB}).  We may 
set $\langle \varphi ^{\dnA }\rangle =0$.  The one-point function $\langle \varphi ^{\upA }(\bbox{x})\rangle $ is essentially Abelian, and 
the Cauchy-Riemann equation for (\ref{ClassGenerB}) yields precisely the same 
soliton equation (\ref{SolitEqVorte}).  The topological charge density is given by 
(\ref{VorteCharg}) with an analytic function $\omega (z)=\omega ^{\upA }(z)$.  The U(1) excitation is 
the vortex.  The Coulomb energy of the vortex excitation is given by 
(\ref{CouloEnerg}) with $\gamma \simeq 0.39$.  There is no antivortex excitation.  Instead of 
it an electron is placed into the spin-down state, as would increase the 
Coulomb energy of the same order as the vortex excitation and the Zeeman 
energy.
\subsection{Skyrmion Excitations}

The SU(2) excitation is generated on the spin-polarized ground state 
(\ref{VacuuMono}) when $\partial _{k}\chi (\bbox{x})=0$ and $\partial _{k}n^{\alpha }(\bbox{x})\not=0$ in (\ref{ClassGenerB}).  The CP 
field is
\begin{equation}
n^{\alpha }(\bbox{x}) = {\omega ^{\alpha }(z)\over \sqrt {|\omega ^{\upA }(z)|^{2}+|\omega ^{\dnA }(z)|^{2}}}, 
\label{GenerSkyrm}
\end{equation}
yielding the wave function
\begin{equation}
\Psi _{\text{Skyrmion}}[\bbox{x}] = 
\prod _{r}\bpmatrix \omega ^{\upA }(z_{r})\\ \omega ^{\dnA }(z_{r})\epmatrix_{\kern-1pt r}\Laugh .
\label{SkyrmWave}
\end{equation}
A simplest choice is given by
\begin{equation}
\Psi _{\text{Skyrmion}}[\bbox{x}] = \bpmatrix z^{q}\\ (\kappa /2)^{q}\epmatrix\Laugh ,
\label{SkyrmOmega}
\end{equation}
with a positive integer $q$, which describes a classical Skyrmion with scale 
$\kappa $ sitting at the origin of the system.  It is clear in (\ref{SkyrmOmega}) that 
the Skyrmion is reduced to the vortex in the limit $\kappa \rightarrow 0$, where there is no 
distinction between the U(1) and SU(2) excitations.  Because the vortex is 
regarded as the small limit of the skyrmion, we do not make a clear 
distinction between them in the QH ferromagnet. 

For the Skyrmion (\ref{SkyrmOmega}) the classical sigma field (\ref{ClassPS}) is 
calculated as
\begin{equation}
s^{x} = \sqrt {1-(s^{z})^{2}}\cos(q\theta ),\quad
s^{y} =-\sqrt {1-(s^{z})^{2}}\sin(q\theta ),\quad
s^{z} = {r^{2q}-(\ell _{B}\kappa )^{2q}\over r^{2q}+(\ell _{B}\kappa )^{2q}},
\label{SkyrmSpin}
\end{equation}
and the Pontryagin number density (\ref{PontrNumbe}) as
\begin{equation}
Q^{P}(\bbox{x}) = {q^{2}\over \pi } {r^{2q-2}(\ell _{B}\kappa )^{2q}\over [r^{2q}+(\ell _{B}\kappa )^{2q}]^{2}} .
\label{SkyrmPontr}
\end{equation}
The spin flips at the Skyrmion center, $\bbox{s}=(0,0,-1)$ at $r=0$, while the 
spin-polarized ground state is approached away from it, $\bbox{s}=(0,0,1)$ for 
$r\gg \kappa \ell _{B}$.  

A Skyrmion excitation modulates not only the SU(2) part but also the 
U(1) part via the relation (\ref{ClassGenerB}).  The Cauchy-Riemann equation reads
\begin{equation}
\partial _{j}\bigl(\a(\bbox{x}) - \ln \sqrt {\rho _{0}+\varrho  (\bbox{x})}\bigr) = -\varepsilon _{jk}K_{k} ,
\label{SkyrmSPNf}
\end{equation}
where
\begin{equation}
K_{k}= -\i\sum _{\alpha  }n^{\alpha *}\partial _{k}n^{\alpha } .
\end{equation}
From (\ref{SkyrmSPNf}) the same soliton equation as (\ref{SolitEqVorte}) is derived,
\begin{equation}
{\nu \over 4\pi }\bbox{\nabla }^{2}\ln\biggl(1+{\varrho  (\bbox{x})\over \rho _{0}}\biggr) - \varrho  (\bbox{x}) = \nu Q^{P}(\bbox{x}) ,
\label{SolitEqSkyrm}
\end{equation}
but the topological charge density now reads
\begin{equation}
Q^{P}(\bbox{x}) = {1\over 2\pi }\varepsilon _{jk}\partial _{j}K_{k}(\bbox{x}) .
\label{SkyrmCharg}
\end{equation}
It is a straightforward calculation \cite{Skyrmion} to show that the charge 
(\ref{SkyrmCharg}) is identical to the Pontryagin number density (\ref{PontrNumbe}).  

The topological charge is evaluated as
\begin{equation}
Q^{P} = \int \dx Q^{P}(\bbox{x}) = {1\over 2\pi i}\oint \text{d}x_{j} K_{j}(\bbox{x}) ,
\label{SkyrmChargA}
\end{equation}
where the loop integration $\oint $ is made to encircle the excitation at infinity 
($|\bbox{x}|\rightarrow \infty $).  The electron number associated with the topological soliton is
\begin{equation}
\Delta N = \int \dx \varrho  (\bbox{x}) = -\nu  Q^{P} .
\label{NumbeElectS}
\end{equation}
The topological charge is determined by the asymptotic value of the CP field 
(\ref{GenerSkyrm}).  We find $Q^{P}=q$ for the Skyrmion (\ref{SkyrmSpin}).  The electron 
number of this soliton is $\Delta N=-\nu q$: It represents the number of electrons 
removed by the Skyrmion excitation.

The Skyrmion spin is given by
\begin{equation}
\Delta N_{\text{s}}= - {1\over 2}\int \dx \bigl\{2S^{z}(\bbox{x})-\rho _{0}\bigr\} = {1\over 2}\int \dx \bigl\{\rho _{0}-\rho (\bbox{x})s^{z}(\bbox{x})\bigr\} .
\label{NumbeFlipp}
\end{equation}
It seems to diverge logarithmically for the Skyrmion (\ref{SkyrmSpin}) with $q=1$.  
This is a fake since the the Zeeman term breaks the spin SU(2) symmetry 
explicitly and introduces a coherent length $\xi _{L}$ into the SU(2) component.  
The Skyrmion configuration (\ref{SkyrmSpin}) is valid only within the coherent 
domain because the coherent behavior of the spin texture is lost outside it.  
By cutting the upper limit of the integration at $r=\kappa \xi _{L}/2$ in (\ref{NumbeFlipp}), 
we obtain
\begin{equation}
\Delta N_{\text{s}} = {\kappa ^{2}\over 2} \ln\biggl({\xi _{L}^{2}\over 4\ell _{B}^{2}}+1\biggr) = {\kappa ^{2}\over 2}\ln\biggl({\sqrt {2\pi }\over 32\widetilde{g}}+1\biggr) ,
\label{NumbeSpinFlip}
\end{equation}
where $\xi _{L}$ is the coherent length given by (\ref{CoherLengt}) and $\widetilde{g}$ is the 
normalized g-factor (\ref{NormaGfacto}).

The density modulation around the Skyrmion is governed by the soliton 
equation (\ref{SolitEqSkyrm}).  This equation has formally the same expression as 
the soliton equation (\ref{SolitEqVorte}) for the vortex excitation.  We may obtain 
an approximate solution in the two limits, the large Skyrmion limit ($\kappa \gg 1$) 
and the small Skyrmion limit ($\kappa \ll 1$).  First, in the large limit we may solve 
(\ref{SolitEqSkyrm}) iteratively as
\begin{equation}
\varrho  (\bbox{x}) = -\nu Q^{P}(\bbox{x}) - {\nu ^{2}\over 8\pi \rho _{0}}\nabla ^{2}Q^{P}(\bbox{x}) + \cdots  .
\label{SolitEqSkyrmA}
\end{equation}
We may approximate it as
\begin{equation}
\varrho  (\bbox{x}) \simeq  -\nu Q^{P}(\bbox{x})={\nu \over \pi } {(\ell _{B}\kappa )^{2}\over [r^{2}+(\ell _{B}\kappa )^{2}]^{2}},\quad \quad \text{for}\quad  \kappa \gg 1 ,
\label{SolitEqSkyrmB}
\end{equation}
for the Skyrmion with $q=1$, where we have used (\ref{SkyrmPontr}).  It agrees with 
the formula due to Sondhi et al. \cite{SkyrmQH}.  On the other hand, the 
topological charge $Q^{P}(\bbox{x})$ is localized in the small limit, $Q^{P}(\bbox{x})\rightarrow q\delta (\bbox{x})$ as 
$\kappa \rightarrow 0$ in (\ref{SkyrmPontr}).  Hence, the solution is given by the vortex 
configuration, 
\begin{equation}
\varrho  (\bbox{x}) \simeq  -\rho _{0}\biggl(1+{\sqrt {2}r\over \ell _{B}}-{r^{2}\over 3\ell _{B}^{2}}\biggr)\e^{-\sqrt {2}r/\ell _{B}} ,\quad \quad \text{for}\quad  \kappa \ll 1 ,
\label{SolitEqSkyrmC}
\end{equation}
which has been derived in (\ref{BetteAppro}).

We evaluate the excitation energy of a Skyrmion with $q=1$ at $\nu =1$.  
It consists of the exchange energy, the electrostatic term and the Zeeman 
term, where the exchange energy (\ref{SpinStiffTerm}) is exactly calculable for one 
Skyrmion,
\begin{equation}
E_{\text{Skyrmion}} = 4\pi \rho _{s}+{e^{2}\over 2\varepsilon }\int \dx\dy {\varrho  (\bbox{x})\varrho  (\bbox{y})\over |\bbox{x}-\bbox{y}|} + g^{*}\mu _{B}B \Delta N_{\text{s}} ,
\label{SkyrmEnergSpin}
\end{equation}
where the Skyrmion spin $\Delta N_{s}$ is given by (\ref{NumbeFlipp}).    It is calculated 
by using (\ref{SolitEqSkyrmB}) and (\ref{NumbeSpinFlip}) for a large Skyrmion, 
\begin{equation}
E_{\text{Skyrmion}} = {e^{2}\over \varepsilon \ell _{B}}\biggl[\sqrt {\pi \over 32}+{\beta \over \kappa }+ {\widetilde{g}\kappa ^{2}\over 2}\ln\biggl({\sqrt {2\pi }\over 32\widetilde{g}}+1\biggr)\biggr],
\label{SkyrmTotalEnerg}
\end{equation}
with $\beta =3\pi ^{2}/64$.  For a sufficiently small Skyrmion, the Coulomb energy is 
calculated with the vortex configuration and is given by (\ref{CouloEnerg}).

The Coulomb energy increases for a smaller Skyrmion while the Zeeman 
energy increases for a larger Skyrmion.  The optimized scale $\kappa $ is obtained 
by minimizing the total energy (\ref{SkyrmTotalEnerg}),
\begin{equation}
\kappa  = \beta ^{1/3}\biggl\{\widetilde{g}\ln\biggl({\sqrt {2\pi }\over 32\widetilde{g}}+1\biggr)\biggr\}^{-1/3} .
\label{OptimScale}
\end{equation}
The Skyrmion excitation energy is given by (\ref{SkyrmTotalEnerg}) together with 
(\ref{OptimScale}).

The parameter $\beta $ describes a strength of the Coulomb energy.  In 
general it is a function of the size $\kappa $, obeying $\beta (\kappa )\rightarrow 0.39\kappa $ as $\kappa \rightarrow 0$ and 
$\beta (\kappa )\rightarrow 3\pi ^{2}/64$ as $\kappa \rightarrow \infty $.  We expand it around an arbitrary point $\kappa _{0}$ as $\beta (\kappa ) 
= \beta _{0}+\beta _{1}(\kappa -\kappa _{0})$.  The $\kappa $-dependence of the Skyrmion energy is the same as in 
(\ref{SkyrmTotalEnerg}) with $\beta =\beta _{0}-\beta _{1}\kappa _{0}$.  The optimized scale is given by 
(\ref{OptimScale}) with the same replacement.  The parameter $\beta $ is calculable only 
if the soliton equation (\ref{SolitEqSkyrm}) is solved for a Skyrmion wih arbitrary 
scale $\kappa $.  Furthermore, there will be a correction to it from a finite 
thickness of the layer.  In this paper we treat $\beta $ as a phenomenological 
parameter.

\begin{figure}[hbt]
\epsfxsize=100mm\epsfbox{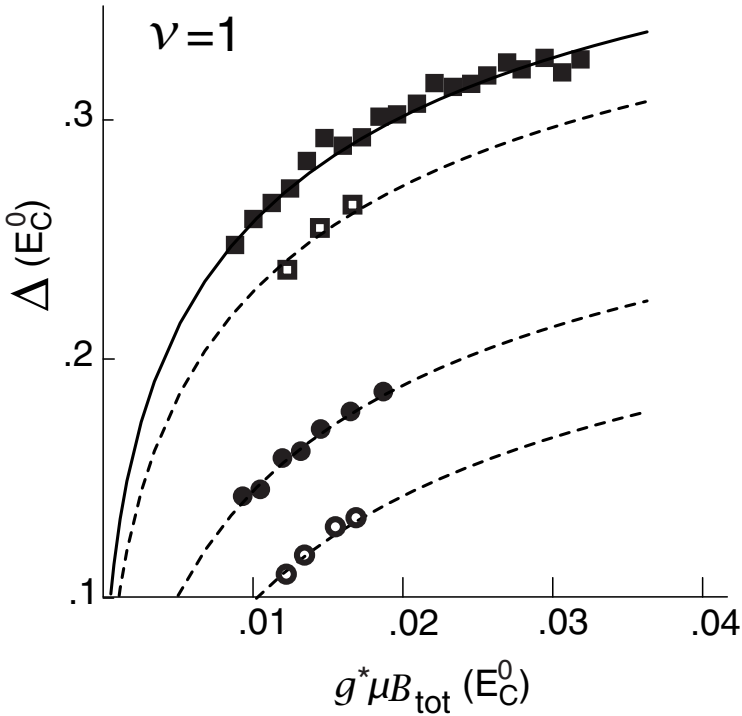}
\renewcommand{\baselinestretch}{0.8}
\caption{\small
Theoretical curves versus experimental data due to Schmeller et al. [11]
for the activation energy in QH ferromagnets.  All the data are fitted 
excellently by the theoretical formula (\ref{SkyrmOffse}) with
(\ref{SkyrmTotalEnerg}),
where an  appropriate offset $\Gamma _{\text{offset}}$ is assumed 
for each sample.
}\label{SkyrEneXPS}
\end{figure}

The creation energy of a Skyrmion-antiSkyrmion pair will be given by 
$2E_{\text{Skyrmion}}$ if they are sufficiently far apart one another.  However, 
it is not this quantity that is observed experimentally.  The activation 
energy $\Delta $ is usually determined by the Arrhenius formula,
\begin{equation}
R_{xx} \propto \exp\bigl[-{\Delta \over 2kT}\bigr],
\end{equation}
where $R_{xx}$ is the magnetoresistivity.  Various factors affect the activation 
energy \cite{FQHEbook}.  For instance, impurities make Coulomb potentials around 
them and lead to a Landau-level broadening.  Phenomenologically, their effects 
result in a subtraction of a certain amount of offset $\Gamma _{\text{offset}}$ from 
the Skyrmion creation energy,
\begin{equation}
\Delta  = 2E_{\text{Skyrmion}} - \Gamma _{\text{offset}} ,
\label{SkyrmOffse}
\end{equation}
where $\Gamma _{\text{offset}}$ increases with disorder.  In FIG.\ref{SkyrEneXPS} we 
have fitted the experimental data due to Schmeller et al. \cite{SkyExpEneA} by the 
formula (\ref{SkyrmTotalEnerg}), where we have used $\beta =0.24$ and an appropriate 
offset $\Gamma _{\text{offset}}$ phenomenologically for each curve.  The theoretical 
curve reproduces all the data remarkably well.  The Skyrmion spin is estimated 
that $\Delta N_{s}\simeq 3.7$ at $\widetilde{g}=0.015$ ($B=3.05$ Tesla).  This estimation is consistent 
with the Hartree-Fock result \cite{SkyrmHartFock} and the experimental result 
based on Knight-shift measurements \cite{SkyExp}. 
\section{Discussions}

We have studied QH ferromagnets based on the improved CB theory.  We 
have investigated excitations confined to the \LLLd.  Charged excitations are 
Skyrmions, carrying the quantized charge $\nu e$ at the filling factor $\nu =1/m$.  
Small Skyrmions are identified with vortices.  We have successfully explained 
the experimental data due to Schmeller et al. \cite{SkyExpEneA}.

We would like to thank A. Sawada, I. Takagi and K. Tsuruse for various 
discussions on the subject.  A partial support is acknowledged from a 
Grant-in-Aid for the Scientific Research from the Ministry of Education, 
Science, Sports and Culture (10138203,10640244).

\end{document}